%% file: main.tex
\pdfoutput=1
\documentclass[pra,aps,twocolumn,10pt,floatfix]{revtex4-2}

\usepackage{graphicx} 
\usepackage{xcolor} 
\usepackage{enumitem}
\usepackage{tikz}
\usepackage{pgfplots}
\pgfplotsset{compat=1.18}
\usepackage{amsfonts}
\usepackage{amsmath,amssymb,amsthm}
\usepackage{graphicx}
\usepackage{fancyhdr}
\usepackage[breakable]{tcolorbox}
\usepackage{bbm}
\usepackage{url}
\usepackage{mathtools}
\usepackage{braket}
\usepackage{upgreek}
\usepackage{dsfont}
\usepackage{collectbox}
\usepackage{braket}
\usepackage{quantikz}
\usepackage{bm}
\usepackage[plain]{fancyref} 
\usepackage[ruled]{algorithm2e}
\usepackage{algpseudocode}
\usepackage{hyperref}
\usepackage{cleveref} 
\usepackage{mfirstuc}

\tikzset{
    vertex/.style={circle, draw=black, fill=blue, inner sep=0pt, minimum size=6pt}
}

\Crefname{figure}{Fig.}{Figs.}
\Crefname{equation}{Eq.}{Eqs.}
\Crefname{section}{Sec.}{Secs.}
\Crefname{appendix}{Appx.}{Appcs.}
\Crefname{algorithm}{Alg.}{Algs.}
\DeclarePairedDelimiter\ceil{\lceil}{\rceil}
\DeclarePairedDelimiter\floor{\lfloor}{\rfloor}
\DeclareMathOperator{\mean}{mean}
\DeclareMathOperator{\sdp}{sdp}

\begin{document}
\title{Iterative quantum optimisation with a warm-started quantum state}

\author{\begingroup
\hypersetup{urlcolor=navyblue}
\href{https://orcid.org/0009-0001-4941-5448}{Haomu Yuan}
\endgroup}
\affiliation{Cavendish Laboratory, Department of Physics, University of Cambridge, Cambridge CB3 0HE, UK}
\email[Haomu Yuan ]{hy374@cam.ac.uk}

\author{Songqinghao Yang}
\affiliation{Cavendish Laboratory, Department of Physics, University of Cambridge, Cambridge CB3 0HE, UK}

\author{Crispin H. W. Barnes}
\affiliation{Cavendish Laboratory, Department of Physics, University of Cambridge, Cambridge CB3 0HE, UK}

\date{\today}

\begin{abstract}
We provide a method to prepare a warm-started quantum state from measurements with an iterative framework to enhance the quantum approximate optimisation algorithm (QAOA). The numerical simulations show the method can effectively address the "stuck issue" of the standard QAOA using a single-string warm-started initial state described in~\cite{Cain2023}. 
When applied to the $3$-regular MaxCut problem, our approach achieves an improved approximation ratio, with a lower bound that iteratively converges toward the best classical algorithms for $p=1$ standard QAOA. Additionally, in the context of the discrete global minimal variance portfolio (DGMVP) model, simulations reveal a more favourable scaling of identifying the global minimal compared to the QAOA standalone, the single-string warm-started QAOA and a classical constrained sampling approach.
\end{abstract}
\maketitle

\section{Introduction} \label{intro}
A good initial state will reduce the convergence time and improve the optimisation quality of the variational quantum algorithm (VQA). Initial state preparation is becoming a crucial pre-processing step for the VQAs. 
For instance, the Hartree-Fock state\cite{Quantum2020,Grimsley2023}, Hardamard transformation~\cite{Farhi2014}, Dicke state~\cite{Cook2020} and warm-started initial states~\cite{Egger2021,Tate2023,yuan2024quantifyingadvantagesapplyingquantum} are widely used in quantum chemistry and classical combinatorial optimisations. In particular, the warm-started method has shown  advantages in many applications. Ref.~\cite{Egger2021} first proposed to use a warm-started initial state for the QAOA by projecting the relaxed quadratic programming results with the $R_y$ rotations. 
Ref.~\cite{Tate2023} claimed a possibility of beating the GW-approximation for the MaxCut by the QAOA utilising the semidefinite programming (SDP) solution with a custom mixing operator in a low-depth circuit.
Ref.~\cite{yuan2024quantifyingadvantagesapplyingquantum} also found a warm-started initial state derived from a rounded SDP solution can improve the QAOA's performance on the DGMVP model when paired with a custom hard-mixing operator in a low-depth circuit. 
Notably, Both Ref.~\cite{yuan2024quantifyingadvantagesapplyingquantum} and Ref.~\cite{Tate2023} showed a potential connection between the warm-started initial state and mixing operator of QAOA across different applications. 
Ref.~\cite{He2023} further found that aligning the initial state with the hard-mixing operator can improve the QAOA's optimisation. However, Ref.~\cite{Cain2023} raised concerns that the optimisation will get stuck with the standard QAOA mixing and cost operator if we start from a good classical string. They found that no states improved after the optimisation as long as the classical string was not sufficiently far away from the optimal solution. Thus, the findings doubt the advantages of using the warm-started method. 

This article proposes a new warm-started initial state preparation method that constructs a quantum state by superposing the best-measured classical strings. We further improve the quantum approximation ratio iteratively using the method. We choose the QAOA as our focus due to its diversified applications and potential efficiency in solving combinatorial optimisation~\cite{Wang2018,Shaydulin2024}. The method can also be expanded to the general VQAs easily. The simulations and analyses on the MaxCut show it can avoid the stuck issue of a single-string warm-started method. The simulations on the DGMVP model show that it can help find a better optimal solution for portfolio optimisation than a standalone QAOA.
The positive results suggest that our method offers a novel perspective for designing quantum algorithms.

The rest of this article is structured as follows. In~\Cref{sec:model}, we review the MaxCut and DGMVP models. In~\Cref{sec:method}, we introduce the iterative warm-started quantum approximate optimisation algorithm, including the QAOA ansatz applied to classical models. In~\Cref{sec:result}, we analyse the simulation results of the algorithm on the MaxCut and the DGMVP models. In~\Cref{sec:discussion}, we summarise the results and discuss the potential directions and applications of the algorithm.

\section{Models}\label{sec:model}
In this section, we introduce two classical combinatorial optimisation models that have been solved by the QAOA and will be applied in this article. 
The first model is the MaxCut problem. An non-loop graph can be written as $G(v,E)$, where $V$ includes the $N$ vertices $\{i\}$ and $E$ includes the set of edges $\{\langle i,j\rangle\}$. 
The MaxCut problem aims to find a binary string $\bm{z} = z_1z_2,\dots,z_n$, corresponding the vertices, such that the total number of edges with different binary values between their vertices is maximised. The objective function is
\begin{equation}
  \underset{\bm{z}}{\operatorname*{argmax}} \ c(\bm{z}) = \sum_{\langle i,j\rangle\in E} c_{\langle i,j\rangle}(\bm{z})\label{eq:cut}
\end{equation}
where $\langle i,j\rangle$ ranges over all edges of a graph $G(V,E)$, $z_i\in\{0,1\}$, and $c_{\langle i,j\rangle}(\bm{z}) = 1$ if $z_i\neq z_j$ and $c_{\langle i,j\rangle}(\bm{z}) = 0$ if $z_i = z_j$. For instance, we often consider a $k$-regular graph, which means each vertex is connected to $k$ edges. A $3$-regular graph, also known as a cubit graph, is a special case of the $k$-regular graph that is widely studied in the QAOA~\cite{Farhi2014,Farhi2020,Wurtz2021,Akshay2021,Larkin2022,Tate2024}.
The MaxCut problem is a well-known NP-hard problem, and we usually use the approximation ratio to evaluate the algorithm's performance. 
Goemans and Williamson provide an approximation ratio of at least $0.8786$ using a technique of randomised rounding the solution to the relaxation of a SDP~\cite{Goemans1995}. An improved approximation ratio of at least $0.9326$ can be achieved when applying the same technique to the $3$-regular graph~\cite{Halperin2004}. The approximation ratio of MaxCut of the classical algorithms is
\begin{align}
 \alpha_c = \frac{\mathds{E}[c]}{c_{\sdp}}\label{eq:alphaapprox_classical}
\end{align} 
where $\mathds{E}[c]$ means the expected value of the cut after randomised rounding, and $c_{\sdp}$ is the optimal solution from the SDP, satisfying $c_{\sdp}\geq c_{\max}$.

\begin{figure}[t]
    \centering
    \includegraphics{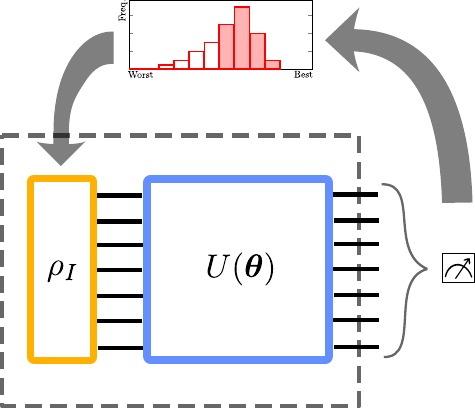}
		\caption{A diagram for iterative warm-started quantum approximation optimisation.}\label{fig:main}
\end{figure}

The other classical combinatorial problem widely concerned by the QAOA is portfolio optimisation~\cite{Hodson2019,Egger2021,Brandhofer2022,Herman2022,He2023,Chen2024,yuan2024quantifyingadvantagesapplyingquantum}. Portfolio optimisation model typically include constraints and requires more complex QAOA ans\"atze than MaxCut problem.
For example, the well-known portfolio model DGMVP has been thoroughly investigated in~\cite{yuan2024quantifyingadvantagesapplyingquantum},
\begin{equation}\label{qdp}
  \hspace*{-1.1cm} 
  \begin{aligned}
      \underset{\bm{w}}{\operatorname*{argmin}} \ & c(\bm{w}) = \bm{w}^\intercal \bm{\Sigma} \bm{w}
  \end{aligned}
\end{equation}
\vspace{-2.5em}
\begin{subequations}
\begin{align}
  \quad \mathrm{s.t.} \quad & \boldsymbol{w}^\intercal \mathbf{1}_n = 1, \tag{4a} \label{qdp:2a}\\
  \quad & {w}_i \in [0,1], \tag{4b} \label{qdp:2b}\\
  \quad & {w}_i/a \in \mathbb{Z}, \ i = 1,2,\dots,n \tag{4c} \label{qdp:2c}
\end{align}
\end{subequations}
where $\boldsymbol{w} = (w_1,w_2,\dots, w_n)^\intercal $ is an $n$-dimensional vector of binarily discretized weights for assets $1, \dots, n$, $\mathbf{1}_n$ is an $n$-dimensional ones vector, and $a$ is the unit trading lot for assets. The $n\times n$ covariance matrix $\boldsymbol{\Sigma}$ and its elements $\sigma_{ij}$ quantify the risk between each pair of assets.

\section{Methods}\label{sec:method}
In this section, we introduce the iterative warm-started quantum approximate optimisation algorithm (see the schematic diagram in~\Cref{fig:main} and the technical details in~\Cref{algorithm:Iterwarm}). The general idea involves iteratively selecting the most successful measurement outcomes from a post-optimisation QAOA ansatz to construct a warm-started initial state, which is then re-optimised using the QAOA. Unlike previous warm-started methods that rely on a single classical string, our approach combines the selected measured strings into a superposed quantum state, offering a more versatile and effective initialisation.

\begin{algorithm}[h]
  \caption{Iterative warm-started quantum approximate optimisation algorithm}
  \label{algorithm:Iterwarm}
  \KwData{$ i =0, \rho^{i}_I; U(\bm\theta ); c: \{0, 1\}^n \rightarrow \mathbb{R};$ $\epsilon, \varepsilon, t > 0;$ $m \in \mathds{Z}^+$.}
  \KwResult{$i,\rho_I^i$.}
  \Repeat{$\|\rho^i_I -\rho^{i-1}_I\| < \epsilon$}{
      $\psi_I = U(\bm\theta )\rho^{i}_I U(\bm\theta )^{\dagger}$\;
      Optimise $\bm\theta$ to $\bm\theta'$ by $c$\;
      $\psi_O = U(\bm\theta' )\rho^{i}_I U(\bm\theta' )^{\dagger}$\;
      $\mathrm{M}(\psi_O) = \{ p_1x_1,p_2x_2,...,p_kx_k\},\sum_{i=1}^{k} p_i = M$\;
      $\tilde\psi^{k} :=  \sum_{j=1}^{k} p'_j |x_j\rangle, p'_j = \sqrt{p_j/m}$\;
      $i+1\leftarrow i $\;
      find a $\rho^i_I$ s.t. $\rho_I^i = \tilde\psi^{t}, t \leq k$\;
  }
\end{algorithm}

In~\Cref{algorithm:Iterwarm}, the initial state $\rho_I^i$ is updated iteratively until it converges under a matrix norm. Each update is referred to as an iteration, with the $i$-th update denoted as Iter-WS($i$). $U(\bm\theta )$ denotes the chosen VQA. 
$\mathrm{M}(\psi_O)$ is an ordered measurement distribution of the state on a complete computational base $\{x_k\}$ with $p_k$ denotes the number of measuring the basis $x_k$. The $c$ is a cost function depending on the models, such as the MaxCut~\Cref{eq:cut} and the DGMVP~\Cref{qdp} models. The $\{x_k\}$ is an ordered set from the best to the worst solutions according to $c$. Thus, $\tilde\psi^{t}$ can guarantee a better approximation ratio once optimising $U(\bm\theta )$ successfully. We name this method of preparing the warm-started quantum state as $t$-order statistic's state preparation method. 

The other method to find $\rho^i_I$ based on the measurements is the percentile method, such as for a $t \in[0,1]$,
\begin{equation}
  \begin{aligned}
\rho^i_I = \sum_{i=1}^{k'} {p'}_{i}|x_i\rangle, \quad {p'}_{i} = \sqrt{\frac{{p'}_i}{\sum_{i=1}^{k'} {p'}_{i}}}
  \end{aligned}
\end{equation}
where, in MaxCut problem, we have
\begin{equation}
  \begin{aligned}
\frac{\sum_{i=1}^{k'} p_{i}x_i }{\sum_{i=1}^{k} p_kx_k}\leq t \leq \frac{\sum_{i=1}^{k'+1} p_{i}x_i}{\sum_{i=1}^{k} p_kx_k}.
  \end{aligned}
\end{equation}

We use the $t$-ordered statistic state in this work as it is easier to fix the depth of the circuit. The percentile method could potentially lead to an exponential increase in the number of included basis state $|x_k\rangle$ and the circuit depth required to prepare the initial state by the qubit size increases. Both methods are interchangeable by modifying $t$ up to a value. The initial state is then prepared using the Permutation Grover-Rudolph method from~\cite{Ramacciotti2024}, which employs an ancillia qubit and multi-controlled gates. Thus, the circuit size of preparing the $t$-ordered statistic state on $N$ qubits' circuit is $N+1$, and the complexity is $O(tN)$. 

In this article, we choose the QAOA's parameterised quantum circuit as the $U(\bm\theta)$. In general, a parameterised quantum circuit of the QAOA can be represented as
\begin{equation}
    U(\bm\theta)\coloneq U(B,\beta_p) U(C,\gamma_p)\dots U(B,\beta_1) U(C,\gamma_1),  \label{eq:qaoaU}
\end{equation}
where $\bm\theta:= (\gamma_1,\beta_1,\dots,\gamma_p,\beta_p)$, $p$ is the layer of the QAOA, and $U(C,\gamma)$ and $U(B,\beta)$ are the cost and mixing operators, respectively. 

For the MaxCut problem, we say a standard QAOA choose the cost and mixing operators as
\begin{align}
    C &= \sum_{\langle i,j\rangle \in E} \frac{(\mathds{1} - Z_iZ_j)}{2}, \ U(C,\gamma) = e^{-i\gamma C},\label{eq:stdqaoaC}\\
    B &=\sum_{j=1}^n X_j, \ U(B,\beta) = e^{-i\beta B},\label{eq:stdqaoaB}
\end{align} 
where $Z, X$ are the Pauli-Z and Pauli-X operators, $\langle i,j\rangle$ ranges over all edges of a graph $G(V,E)$, and $\gamma,\beta \in [0,2\pi]$. For the $i=0$ situation, the initial state is a Hadamard  transformation as
\begin{align}
  \rho_I^{0}=\operatorname{H}^{\otimes n}|0\rangle^{\otimes n}, \label{Eq:iniHstate} 
\end{align}
where $\operatorname{H}$ is a Hardamard gate. 

For the DGMVP model, we choose the cost and nearest-neighbour mixing operator, with each asset binary encoded with $l$-length qubits' measurement variables. The cost operator is given by 
\begin{equation}\label{eq:DGMVP_hamiltonian}
	\begin{aligned}
  		C &=\frac{a^2}{2}\sum_{i<j}\sum_{k_1,k_2} \sigma_{ij}b^{k_1} b^{k_2} Z^{k_1}_{i} Z^{k_2}_{j} \\
      & + \frac{a^2}{4} \sum_{i}\sum_{k_1\neq k_2}\sigma_{ii}b^{k_1} b^{k_2} Z^{k_1}_{i} Z^{k_2}_{i} \\
      & - \frac{a}{2} \sum_{i} \sum_{k}\tilde{\sigma}_{i} b^{k} Z^{k}_{i}, \\
      U&(C,\gamma) = e^{-i\gamma C},
	\end{aligned}
\end{equation}
where $Z_i^k,Z_i^{k_1},Z_i^{k_2},$ and $Z_j^{k_2}$ are Pauli-Z operators, $b^{k} = 2^{k-1} a$. We drop a constant offset as it contributes to a global phase. The mixing operator is 
\begin{equation}\label{eq:DGMVP_nnmixing}
	\begin{aligned}
		{U}(B,\beta) =  \prod_{t-t'=1} \tilde{S}_{tt'}(\beta) \tilde{P}^e_{tt'}(\beta) \tilde{P}^o_{tt'}(\beta) \tilde{S}_{tt'}(\beta), 
	\end{aligned}
\end{equation} 
where
\begin{equation}\label{eq:DGMVP_nnmixing_sub1}
	\begin{aligned}
		 & \tilde{S}_{tt'}(\beta) = \bigotimes_{k=1}^l e^{\beta S^k_{tt'}},      \\
		 & \tilde{P}^e_{tt'}(\beta) = \bigotimes_{k \in \mathcal{K}_1} e^{\beta P^{k}_{tt'}}, \\
		 & \tilde{P}^o_{tt'}(\beta) = \bigotimes_{k \in \mathcal{K}_2 } e^{\beta P^{k}_{tt'}}, 
	\end{aligned}
\end{equation}
and 
\begin{equation}\label{eq:DGMVP_nnmixing_sub2}
	\begin{aligned}
     & S_{tt'}^k \coloneqq Q^{k\dagger }_{t}Q^k_{t'} - Q^k_{t}Q^{k\dagger}_{t'}, \\
     & P_{tt'}^{k} \coloneqq Q^{{k+1}\dagger}_{t} Q^{k}_{t} Q^{k}_{t'}- Q^{{k+1}}_{t} Q^{k\dagger}_{t} Q^{k\dagger}_{t'}, \\
		 & \mathcal{K}_1 = \left\{2c_1 \mid c_1 \in \left[1, \floor*{l/2}\right] \cap \mathbb{Z}\right\}, \\
     & \mathcal{K}_2 = \left\{2c_2-1\mid \ c_2 \in \left[1, \ceil*{l/2}\right] \cap \mathbb{Z} \right\},
	\end{aligned}
\end{equation}
where the values of $t-t'$ reduced modulo $n$, $\mathcal{K}_1, \mathcal{K}_2$ reduced modulo $l$, $S_{tt'}^k $ and $P_{tt'}^{k}$ are two and three-qubit excitation operators~\cite{Yordanov2020,yuan2024quantifyingadvantagesapplyingquantum}, and $Q^{k}_{t}, Q^{k}_{t'},$ and $Q^{{k+1}}_{t}$ follows Jordan-Wigner encoding~\cite{Jordan1928} as follows:
\begin{equation}\label{qubitcreaanni}
	\begin{aligned}
		Q^{k\dagger }_{t} = \frac{1}{2}(X^k_t-i Y^k_t), \quad Q^k_t = \frac{1}{2}(X^k_t+i Y^k_t),
	\end{aligned}
\end{equation}
where the super- and subscript indicates the functioned qubits. The nearest-neighbour mixing operator will limit the measurement space under the fix capital constraint~[\Cref{qdp:2a}] and is considered as a hard mixing operator~\cite{Hadfield2019}. The initial state is chosen as the maxbias initial state as:
\begin{equation}\label{eq:maxbias}
    \rho_I^0 =X^{\otimes l}_t |0\rangle^{\otimes n}
\end{equation}
where $X^{\otimes l}_t$ is a tensor product of $l$ Pauli-X operators functioned on the $l$-qubit encoded variable $w_t$, and $\mathds{1}$ is an identity operator. 

\section{Results}\label{sec:result}
In this section, we conduct simulations using the standard QAOA ansatz[\Cref{eq:stdqaoaC,eq:stdqaoaB}] to solve the MaxCut problem on the $3$-regular graph[\Cref{eq:cut}], as well as simulation of employing a QAOA ansatz[\Cref{eq:DGMVP_hamiltonian,eq:DGMVP_nnmixing}] to solve the DGMVP model. 
In the simulations, we randomly generate the $3$-regular graphs and DGMVP instances and apply \Cref{algorithm:Iterwarm}, utilising the $t$-order statistic's state concatenating with corresponding QAOA ansatz. We consider stochastic noise in the simulations for shot measurements, quantified by two parameters: $m$, representing the number of measurement shots of estimating expectation values during the circuit optimisation, and $M$, specifying the number of shots of estimating the expectation value of a post-optimised quantum circuit. 
The classical optimiser is chosen as Dual Anealing(DA), with the maximum number of expectation values evaluations allowed during optimisation denoted as $\mathcal{I}$. 

To benchmark the performance of QAOA's approximation, we choose a normalised approximation ratio as suggested in~\cite{Zhou2022}:
\begin{align}
 r = 1- \frac{\langle C\rangle_{M}}{c_{\max}},\label{eq:alphaapprox}
\end{align} 
where $\langle C\rangle_{M} \coloneqq \langle\bm{\gamma},\bm{\beta}|C|\bm{\gamma},\bm{\beta}\rangle_{M}=\frac{1}{k}{\sum_{i = 1}^{M} p_i c(x_i)}$, $c_{\max}$ is a bruteforced MaxCut value, and $\frac{\langle C\rangle_{M}}{c_{\max}}$ is MaxCut approximation ratio. 
To assess the improvement in each iteration, we define the relative change ratio as 
\begin{align}
 R = \frac{r_{\rho_I}-r}{r_{\rho_I}},
\end{align} 
where $r_{\rho_I}$ is the approximation ratio of prepared initial state, and $r$ is the approxiamtion ratio of the post-optimsied circuit. To evaluate the convergence rate of optimisation, we calculate the percentage of state that cannot be further optimised at each step from a stochastic analysis: 
\begin{align}
 P = \frac{\mathrm{N}_{total}- \mathrm{N}_{static}}{\mathrm{N}_{total}}
\end{align} 
where $\mathrm{N}_{static}$ is the number of states that converges to a static state cannot be improved under an error tolerance $\epsilon$, and $\mathrm{N}_{total}$ is the total number of generated instances. It is noted that $N_{static}$ will increase cumulatively as warm-started iterations progress.

The mean value approximation ratio of the DGMVP model using QAOA is estimated as
\begin{equation}\label{eq:dgmvp_appr_mean}
 \begin{aligned}
 \alpha_{\mean} = \frac{\langle C\rangle_{M} - f_{\min} }{f_{\max} - f_{\min}},
 \end{aligned}
\end{equation}
where $\langle C \rangle_{M}$ is specified in~\Cref{eq:alphaapprox}, $f_{\min}$ and $f_{\max}$ are the global minimal and maximal values of the cost function~\Cref{qdp}, respectively. 

Additionally, we use the minimal value approximation ratio for the DGMVP model, defined as
\begin{equation}\label{eq:appr_min}
 \begin{aligned}
 \alpha_{\min} = \frac{ \min_{i\in\left[1,M\right]}\{ \langle \bm{w}_i|C|\bm{w}_i \rangle \}- f_{\min} }{f_{\max} - f_{\min}},
 \end{aligned}
\end{equation}
where the $\bm{w}_i$ is the $i$-th measured weight vector. The probability of measuring the minimal value, $\min_{i\in\left[1,M\right]}\{ \langle \bm{w}_i|C|\bm{w}_i \rangle\}$, is denoted as $\mathrm{P}_{\min}$, and the probability of measuring the global minimal value (the brute-forced optimal solution) is denoted as $\mathrm{P}_{gm}$. To benchmark the scaling of $\mathrm{P}_{gm}$, we use a classical constrained random sampling method, where the probability of measuring the optimal solution is given by~\cite{yuan2024quantifyingadvantagesapplyingquantum},
\begin{equation}\label{eq:ccrs}
 \begin{aligned}
   \mathrm{P}_c(n,l) = \frac{1}{B(n,l)}, 
 \end{aligned}
\end{equation}
where $B(n,l)= \binom{2^l+n-2}{n-1}$, $n$ is the number of assets, and $l$ is the length of binary variables.

\begin{figure}[h]
  \centering
    \includegraphics{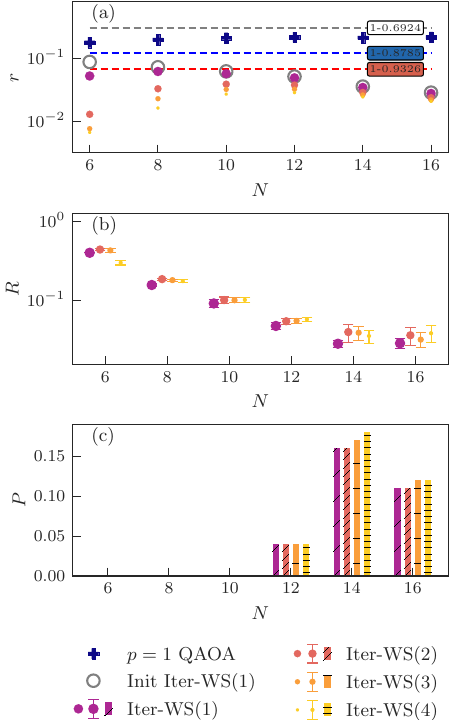}
    \caption{The $r$(a), $R$(b) and $P$(c) for the iterative warm-started quantum approximate optimisation applied over four iterations to the $100$ randomly generated $3$-regular graphs, as a function of graph size $N$. The quantum ansatz employs a $p=1$ standard QAOA with a $20$-th ordered statistic state. The classical optimiser is DA with  $\mathcal{I} = 5000$, $m = 8000$, and $M = 8000$.}\label{fig:scaling_p1}
\end{figure}

\begin{figure}[h]
  \centering
    \includegraphics{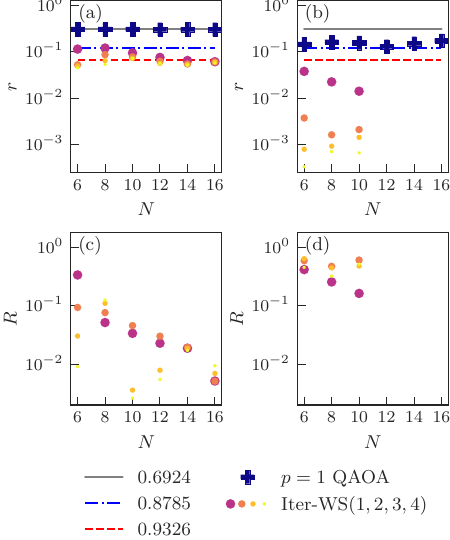}
    \caption{The worst-cases(a)(c) and best-cases(b)(d) of approximation ratios $r$ for the iterative warm-started quantum approximate optimisation in each iteration applied over four iterations to $3$-regular graphs, as a function of graph size $N$. The hidden dots is (b)(d) mean the instances with $r=0$. The quantum ansatz employs a $p=1$ standard QAOA with a $20$-th ordered statistic state. The classical optimiser is DA with $\mathcal{I} = 5000$, $m = 8000$, and $M = 8000$. }\label{fig:scaling_p1_worst_comp}
\end{figure}

\begin{figure}[h]
 \centering \includegraphics{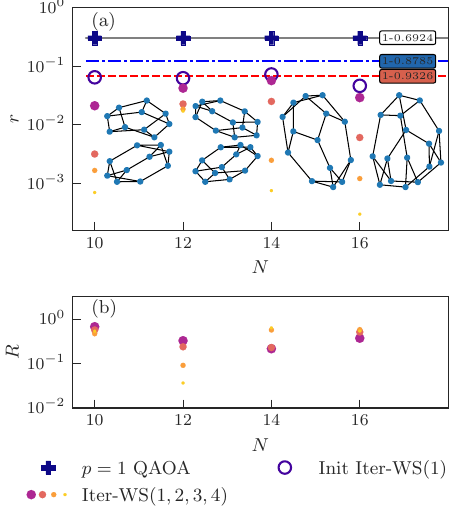}
 \caption{The $r$(a) and $R$(b) for the iterative warm-started quantum approximate optimisation applied over four iterations to the non-planar worst-case of the $3$-regular graphs for a $p=1$ standard QAOA, as a function of graph size $N$. The quantum ansatz employs a $p=1$ standard QAOA with a $20$-th ordered statistic state. The worst-case graphs are displayed next to their corresponding data points. The classical optimiser is DA with $\mathcal{I} = 5000$, $m = 8000$, and $M = 8000$. 
 }\label{fig:scaling_p1_worst}
\end{figure}

In~\Cref{fig:scaling_p1,fig:scaling_p1_worst_comp,fig:scaling_p1_worst}, we apply the iterative warm-started method to improve the $p=1$ standard QAOA's optimisation for the $3$-regular graph.
To evaluate the optimisability of the prepared warm-started quantum state, we set $U(\bm \theta)$ with $\bm\theta = 0$ and assess the results using relative change ratio ($R$), and convergence rate ($P$). The simulation results are summarised in~\Cref{fig:scaling_p1}, with the best-case and worst-case are detailed in~\Cref{fig:scaling_p1_worst_comp}. We find $r$ continuously decreases, and $R$ remains positive and stable, as increasing iterations. The low value of $P$ (below 20\%) indicates the majority of graphs are successfully optimised. These findings suggest that the high optimisanility of the generated warm-started quantum state. 
Interestingly, the iterative warm-started optimisation approach shows significant improvement for smaller graphs $(N\leq 10)$. However, both $R$ and $P$ become worse as $N$ increases, potentially due to the influence of increased stochastic noise on classical optimisers when using the same measurement shots for the larger graph. This  phenomenon aligns with the prior discussions on the effect of stochastic noise on optimisation~\cite{yuan2024quantifyingadvantagesapplyingquantum}.  In addition, these experimental findings aligns with our refined arguments based on the theoretical work first proposed in Ref.\cite{Cain2023}. We show in Appendix \ref{thermodynamic} and Appendix \ref{compression} that iterative post-selection helps resolve the stuck issue. For example, adaptively selecting the good initial state is equivalent to reshaping the thermodynamic equilibrium landscape by energy injection. However, the new search space for the good string suffers from the scaling problem as mentioned above--as the graph size grows, more iterations are required to keep the same effect of improvements.
Despite this, we observe a decreasing trend in $r$ as $N$ increases, suggesting that the iterative warm-started method may be more effective in finding optimal solutions for larger graphs. Notably, the worst-case $r$ values converge toward the approximation ratio of the best classical algorithm, $r= 1 - 0.9326$ after several iterations, underscoring the method's potential for surpassing classical benchmarks.

To further investigate the improvement of the iterative warm-started method to the standard QAOA, we design the worst-case $3$-regular graph for the $p=1$ QAOA optimisation---see~\Cref{fig:scaling_p1_worst}. These graphs contain the specific subgraph $g_6$ at every edge, which has been shown to impose a bounded $r = 1-0.6924$ on $p=1$ standard QAOA---see ~\Cref{sec:specialworst}. Furthermore, we focus on nonplanar graphs, as in~\cite{Harrigan2021}, due to the lack of simple classical methods to solve their MaxCut---see~\Cref{sec:specialworst} for more discussions on a planar graph. To exclude simpler cases, the designed graphs exclusively contain $g_6$ subgraph and rule out the planar configurations. The simulation confirms that the $p=1$ standard QAOA is indeed constrained by the theoretical bound of $r = 1-0.6924$. However,  applying multiple iterations of~\Cref{algorithm:Iterwarm} improves $r$ to values surpassing the best classical algorithms for $3$-regular graphs. Additionally, $R$ remains almost uninfluenced by the graph's size, which indicates our method's robustness.

\begin{figure}[t]
    \centering
      \includegraphics{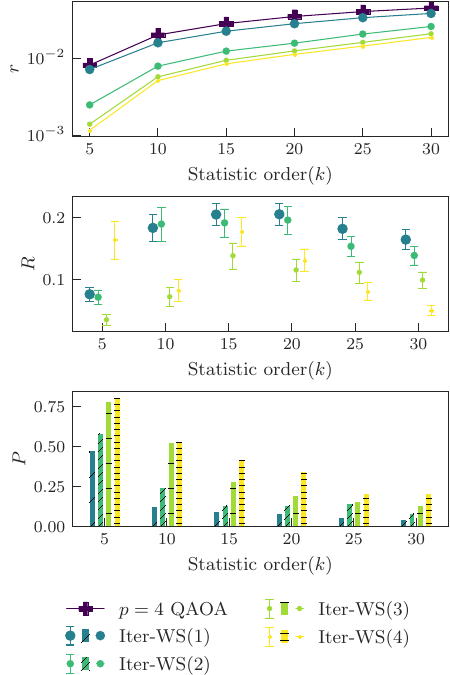}
      \caption{The $r$(a), $R$(b) and $P$(c) for the iterative warm-started quantum approximate optimisation applied over four iterations to $100$ randomly generated $3$-regular graphs with $12$ vertices each, as a function of the prepared initial statistic state order $k$. The quantum ansatz employs a $p=4$ standard QAOA. The classical optimiser is DA with $\mathcal{I} = 5000$, $m = 8000$, and $M = 8000$.}\label{fig:p_4_scaling_percentile}
\end{figure}

In~\Cref{fig:p_4_scaling_percentile}, we analyse the impact of the order $t$ of prepared warm-started initial state to the iterative warm-started optimisation using a $p=4$ standard QAOA. The simulations reveal that $r$ improves with increasing iterations. We observe the most significant $R$ occurs between the $15$- and $20$-th statistic ordered initial state during the first two iterations.

\begin{figure}[h]
  \centering
    \includegraphics{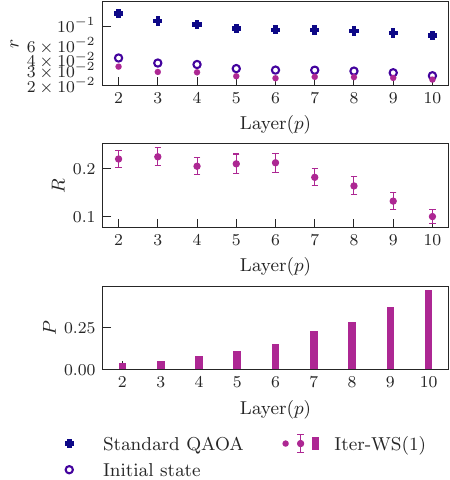}
    \caption{The $r$(a), $R$(b) and $P$(c) for the iterative warm-started quantum approximate optimisation applied over one iteration to $100$ randomly generated $3$-regular graphs with $12$ vertices each, as a function of the layer $p$ of the standard QAOA. The prepared initial state is a $20$-th ordered statistic state. The classical optimiser is DA with $\mathcal{I} = 5000$, $m = 8000$, and $M = 8000$. }\label{fig:p_list_20}
\end{figure}

In~\Cref{fig:p_list_20}, we examine the effect of increasing the standard QAOA layer $p$ on the optimisation of the warm-started quantum initial state. The results show that $r$ will continue to improve as $p$ increases[see~\Cref{fig:p_list_20}(a)]. For the relative change ratio $R$, we observe an initial plateau in the small $p$ region, following a steady decline beyond a certain point [see~\Cref{fig:p_list_20}(b)]. Additionally, $P$ increases steadily as increasing $p$[see~\Cref{fig:p_list_20}(c)]. The trend of more failure could stem from the fixed classical optimisation step (including the optimser and its parameters) used to optimise parameterised ans\"atz with varying numbers of parameters. A similar saturation of $r$ is observed in the standard QAOA optimisation at large $p$ values[see~\Cref{fig:p_list_20}(a)]. Further discussion on the impact of QAOA layers, considering the order of the prepared initial state, is provided in~\Cref{sec:app_p_impact}.

\begin{figure}[htbp]
  \centering
    \includegraphics{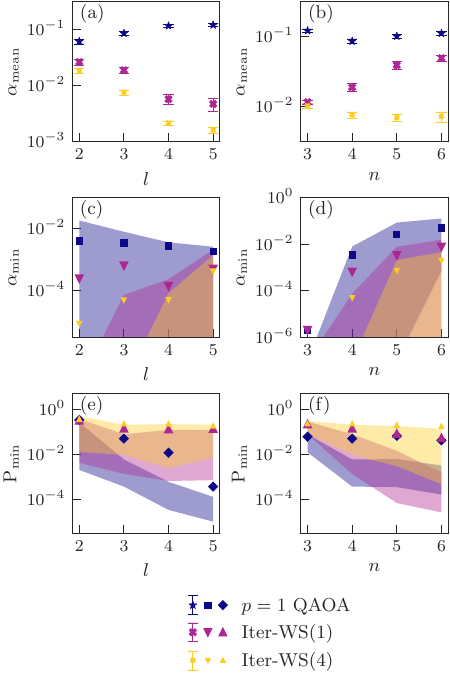}
    \caption{The performance of the iterative warm-started quantum approximate optimisation on the $100$ randomly generated DGMVP instances using a $p=1$ QAOA with a nearest-neighbour mixing operator, as a function of $l$ with $n=4$ (the left three panels) and $n$ with $l=3$ (the right three panels): $\alpha_{\mean}$(a)(b), $\alpha_{\min}$(c)(d), and $\mathrm{P}_{\min}$(e)(f). The initial state is prepared using a $20$-th ordered statistic state. Error bars in (a)(b) represent the standard error. Shaded regions in (c)(d) indicate 10\%-90\% percentiles of the data, and shaded regions in (e)(f) indicate 30\%-70\% percentiles of the data. Hidden dots in (c) correspond to the instance with $\alpha_{\mean} = 0$.
    The classical optimiser is DA with $\mathcal{I} = 2000$,  $m = 16$, and $M = 2^{18}$. }\label{fig:dgmvp_p=1_meanmin}
\end{figure}

\begin{figure}[htbp]
  \centering
    \includegraphics{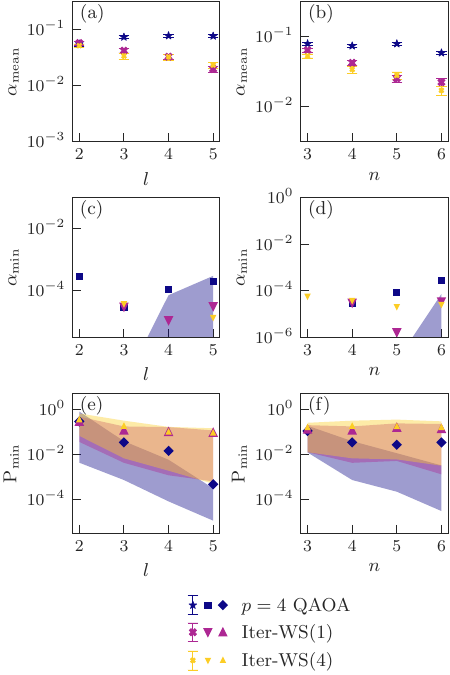}
    \caption{The performance of the iterative warm-started quantum approximate optimisation on the $100$ randomly generated DGMVP instances using a $p=4$ QAOA with a nearest-neighbour mixing operator, as a function of $l$ with $n=4$ (the left three panels) and $n$ with $l=3$ (the right three panels): $\alpha_{\mean}$(a)(b), $\alpha_{\min}$(c)(d), and $\mathrm{P}_{\min}$(e)(f). The initial state is prepared using a $20$-th ordered statistic state. Error bars in (a)(b) represent the standard error. Shaded regions in (c)(d) are 10\%-90\% percentiles of the data, and shaded regions in (e)(f) are 30\%-70\% percentiles of the data.  Hidden dots in (c)(d) correspond to the instances with $\alpha_{\mean} = 0$.
    The classical optimiser is DA with $\mathcal{I} = 2000$, $m = 16$, and $M = 2^{18}$. }\label{fig:dgmvp_p=4_meanmin}
\end{figure}

\begin{figure}[t]
  \centering
    \includegraphics{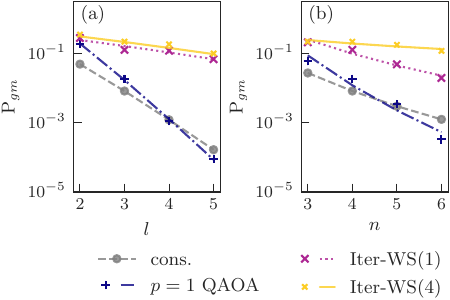}\\
    \vspace{-1cm}
    \includegraphics{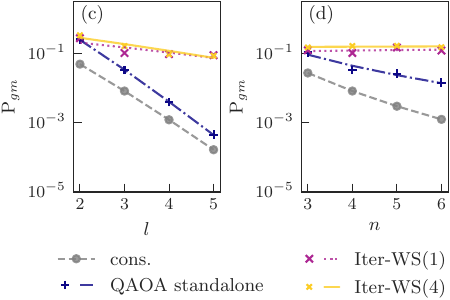}
    \caption{Comparison of the ability to locate DGMVP solutions using the iterative warm-started quantum approximate optimisation applied for four iterations versus a classical constrained sampler for the $100$ randomly generated DGMVP instances, as a function of $l$ with $n=4$ (a)(c) and $n$ with $l=3$ (b)(d). The quantum ansatz employs a $p=1$ QAOA(top two panels) and $p=4$ QAOA(bottom two panels) with a nearest-neighbour mixing operator and a $5$-th ordered statistic state.
    The classical optimiser is DA with $\mathcal{I} = 2000$, $m = 16$, an $M = 2^{18}$.
    Each plus and cross is the average of the measured data, and the lines of the same colour fit the cross points using $ \bar{\mathrm{P}}_{gm} = a \cdot P_c(n,l)^b$ where $P_c(n,l)$ is the probability of measuring the optimal solution with a classical constrained sampler. }\label{fig:dgmvp_dgmvp}
\end{figure}

In~\Cref{fig:dgmvp_p=1_meanmin,fig:dgmvp_p=4_meanmin,fig:dgmvp_dgmvp}, we evaluate the performance of the iterative warm-started optimisation method applied to portfolio optimisation. To ensure a general assessment of our method's applicability, we initialise the circuit parameters $\theta$ randomly. Thus, the comparison of $R$ and $P$ are omitted from the DGMVP simulation due to the the classical optimiser's limited ability to consistently identify the same global optimal parameters.

In~\Cref{fig:dgmvp_p=1_meanmin}(a)(b), we observe that $\alpha_{\mean}$ decreases significantly as increasing iterations compared to the $p=1$ QAOA standalone. Additionally, the iterative warm-started method improves $\alpha_{\min}$ to a lower value [see ~\Cref{fig:dgmvp_p=1_meanmin}(c)(d)] and enhance $\mathrm{P}_{\min}$ to a higher value [see ~\Cref{fig:dgmvp_p=1_meanmin}(e)(f)] as iterations progress. Notably, after several iterations, the results show a reduced $\alpha_{\mean}$ and a stable $\mathrm{P}_{\min}$ across different $l$ and $n$, highlighting the method's robustness to the stochastic noise. 

In~\Cref{fig:dgmvp_p=4_meanmin}, we observe the iterative warm-started method also improves the $\alpha_{\mean}$, $\alpha_{\min}$ and $\mathrm{P}_{\min}$ for $p=4$ QAOA standalone. However, the further iterations provide limited additional improvements. This might indicate that the performance of the iterative warm-started method is constrained by the classical optimiser's effectiveness when applied to deeper QAOA circuits. A similar phenomenon has been discussed in the standard QAOA's $p$ layer simulations. 
Importantly, our method achieves $\alpha_{\min}$ values approaching $0$ while maintaining consistent $\mathrm{P}_{\min}$ across all studied $n$ and $l$ configurations---see~\Cref{fig:dgmvp_p=4_meanmin}(c)(d)(e)(f). This highlights the efficiency and stability of the iterative warm-started approach for identifying low-risk portfolios. 

To further investigate the method in finding DGMVP solutions, we compared the probabilities of generating the DGMVP solutions, $P_{gm}$, from the post-optimised quantum circuit to those from a classical constrained uniform sampler, $P_c(n,l)$[see~\Cref{eq:ccrs}]. This comparison was analysed as a function of  $l$ and $n$---see \Cref{fig:dgmvp_dgmvp}. We fit the mean of $P_{gm}$ using the function $\bar{\mathrm{P}}_{gm} = a \cdot P_c(n,l)^b$, where the inverse ratio of the exponents, $-1/b$, reflects a power-law scaling relative to the classical constrained sampler.
For $l$-scaling, we find $b =-1.36\pm 0.00$ for $p=1$ QAOA standalone optimisation, $b =-0.22\pm 0.00$ after the first iteration of the iterative warm-started optimisation, and $b=-0.21\pm 0.00$ after the fourth iteration---see~\Cref{fig:dgmvp_dgmvp}(a). We also find $b =-1.11\pm 0.00$ for $p=4$ QAOA standalone optimisation, $b =-0.17\pm 0.01$ after the first iteration of the iterative warm-started optimisation, and $b=-0.23\pm 0.00$ after the fourth iteration---see~\Cref{fig:dgmvp_dgmvp}(c). These results indicate that our quantum algorithm achieves a more favourable power-law scaling with $l$ compared to the constrained classical sampling method, although the improvement is limited when increasing iterations. 

For $n$-scaling, we find $b =-1.65\pm 0.07$ for $p=1$ QAOA standalone optimisation, $b =-0.78\pm 0.01$ after the first iteration of the iterative warm-started optimisation, and $b=-0.19\pm 0.01$ after the fourth iteration---see~\Cref{fig:dgmvp_dgmvp}(b). These results highlight a significant improvement in scaling with $n$ as iterations increase for $p=1$ QAOA. We also find $b =-0.61\pm 0.01$ for $p=1$ QAOA standalone optimisation, $b =0.03\pm 0.01$ after the first iteration of the iterative warm-started optimisation, and $b=0.01\pm 0.00$ after the fourth iteration---see~\Cref{fig:dgmvp_dgmvp}(d). The positive $b$ values demonstrate a much better power-law scaling and robustness to stochastic noise. Overall, the scaling with $n$ achieved by our method is more favourable than that of the constrained classical sampling method.

Thus, the scalings in~\Cref{fig:dgmvp_dgmvp} demonstrate demonstrate that our quantum algorithm may asymptotically reduce the number of samples required to identify the DGMVP solution compared to the classical constrained sampler. Moreover, our method achieves more favourable scaling than the layerwise $p=4$ QAOA using single bit string warm-started method reported in~\cite{yuan2024quantifyingadvantagesapplyingquantum}, where the converted $b$ values were $b=-0.76\pm0.3$ for $l$-scaling and $b=-0.66 \pm 0.01$ for $n$-scaling. 

\section{Conclusions}\label{sec:discussion}
In this article, we have provided an iterative optimisation method with warm-started quantum state preparation to enhance the performance of QAOA. Further, we analysed the effectiveness of the method in solving the MaxCut and DGMVP models through numerical simulations. The results demonstrate that the warm-started quantum state significantly improves the approximation ratio for solving $3$-regular graphs using standard QAOA and remains optimisable over multiple iterations. Additionally, for the DGMVP model, our method enhances the mean value approximation ratio, minimum value approximation ratio, and probability of measuring the minimum value. Furthermore, fitting the probability of measuring the global minimal also suggests our quantum algorithm may asymptotically outperform the classical method and the previous warm-started method using a single bit string, requiring fewer samples to find the DGMVP solutions in both $n$ and $l$ scaling. Additionally, the stable performance across various $n$ and $l$ configurations highlights the robustness of the method to stochastic noise. 

However, we observe that the improvement plateaus as iterations increase. The iterative enhancement is also more pronounced in low-depth QAOA circuits, likely due to limitations in classical optimisation when handling circuits with a larger number of parameters.

In short, we showed that the proposed warm-started method enables iterative improvements in QAOA for solving MaxCut and DGMVP problems. Our analyses confirmed that the prepared warm-started initial quantum state will not get stuck in the optimisation. Further works will explore alternative other warm-started initial quantum states, such as those reconstructed via the tomography, establish theoretical bounds on the approximation ratio, and extend the method to other VQAs. 

\section{Acknowledgements}
The authors thank Christopher K. Long for valuable discussions.
\bibliographystyle{unsrt}
\bibliography{main}
\newpage
\appendix

\section{Energy injection for the thermodynamic Argument}\label{thermodynamic}

The original thermodynamic argument presented in \cite{Cain2023} suggests that warm-started QAOA remains stuck due to the assumption that the initial warm-started bitstring behaves as a sample from a thermal ensemble. This assumption leads to an upper bound on the expected cost function value improvement. Specifically, it is argued that for a warm-started string $w$, the improvement is limited by the thermality coefficient $\varepsilon_w$, which scales as $O(1/\sqrt{m})$. This results in the bound:
\begin{equation}
    \frac{1}{m} \langle w | U_w^\dagger C U_w | w \rangle \leq c(w) + 2\varepsilon_w + 4\delta,
\end{equation}
where $c(w) = C(w)/m$ is the initial cost function per edge, and $\delta$ accounts for corrections due to non-tree-like neighborhoods in the graph. The bound implies that for large systems, the improvement vanishes, preventing significant progress in QAOA optimization.

However, this argument assumes that QAOA is applied only once, without iterative refinement. When iterative warm-starting is introduced, the system no longer behaves as a closed thermal ensemble. Instead, it undergoes a non-Markovian feedback process, where each iteration effectively injects energy into the system, continuously modifying the thermality coefficient. To analyze how the system escapes thermal equilibrium, let us define the improvement at each iteration,
\begin{equation}
    \Delta C_k = C_k - C_{k-1}.
\end{equation}
Ref.\cite{Cain2023} assumes that local thermalization prevents improvement, which would suggest
\begin{equation}
    \Delta C_k \leq O(1/\sqrt{m}).
\end{equation}
However, each step modifies the thermality coefficient within the iterative framework, effectively resetting the local thermal equilibrium. To model this formally, define the accumulated improvement after $K$ iterations
\begin{equation}
    C_k - C_0 = \sum^K_{k=1}\Delta C_k.
\end{equation}
If we denote the thermality coefficient after $K$ iterations as $\varepsilon_{w_K}$, its iterative growth can be approximated by:
\begin{equation}
    \varepsilon_{w_K} = \varepsilon_{w_0} + D K,
\end{equation}
where $D$ is a diffusion constant characterizing the perturbation per iteration. Since the improvement bound depends on $\varepsilon_w$, this leads to an overall improvement scaling as:
\begin{equation}
    C_K - C_0 = O(K/\sqrt{m}).
\end{equation}
Unlike the single-step QAOA case, where improvement is constrained to $O(1/\sqrt{m})$, iterative refinement compounds these gains. For sufficiently large iterations $K$, the improvement fully lifts the bound imposed by the original thermal argument. In practice, since the size of the benchmarking is not too big and the graph size has a quadratic nature with the number of iterations, a few iterative QAOA would suffice to solve the issues. For example, one can choose $K=\sqrt{m}$. This result demonstrates that local equilibrium assumptions do not constrain iterative warm-started QAOA and can systematically escape local minima over multiple iterations. It also explains why the adaptive approach is not as pronounced as it is with small benchmarks.

\section{Relaxing the compression space}\label{compression}

The compression argument in \cite{Cain2023} claims that the probability of improving from an initial bitstring $w$ decreases exponentially with system size when QAOA depth $p$ is sublinear. This argument is based on the ratio of the number of initial low-cost states to the number of high-cost states. Specifically, given $d_0$ initial states at cost $C_0$ and $d_1$ states at cost $C_1$ or higher, the probability of improving to $C_1$ is upper bounded as:
\begin{equation}
    \Pr(\text{improvement}) \leq \left(\frac{2 d_1}{d_0}\right)^{\frac{1}{2p+2}} (16\pi p n^2)^{\frac{p}{p+1}},
\end{equation}
where $n$ is the number of qubits and $p$ is the depth of the circuit. When $d_1 \ll d_0$, this probability shrinks exponentially in $n$, implying that QAOA cannot reliably improve the cost function at sublinear depths.

However, again, this argument assumes that QAOA applies a set of unitary transformations $U(\gamma, \beta)$ \textit{once}. The probability distribution evolves over multiple steps in an iterative scheme, expanding the reachable state space. Denoting $d_k$ as the number of states reachable at step $k$, we define the iterative expansion process:
\begin{equation}
    d_K = d_1 + \sum_{k=2}^{K} g_k d_{k-1},
\end{equation}
where $g_k$ represents an amplification factor per iteration. Approximating this as a continuous growth process leads to:
\begin{equation}
    \frac{dD}{dk} = \lambda D_k,
\end{equation}
which solves to an exponential expansion:
\begin{equation}
    D_K = D_0 e^{\lambda K}.
\end{equation}
Since the original compression bound predicts an exponential suppression $D_1 \leq D_0 e^{-a n}$, iterative QAOA counteracts this effect if $\lambda K \geq a n$, breaking the original bound.

Thus, in an iterative setting, the probability of reaching higher-cost bitstrings increases over time, ultimately approaching unity for sufficiently large $K$. This result directly contradicts the compression argument, demonstrating that iterative warm-started QAOA systematically overcomes the limitations of single-shot QAOA and allows for continued improvement beyond the constraints imposed by the initial state distribution.

\section{Discussions of the worst-case of solving 3-regular graph using p=1 standard QAOA}\label{sec:specialworst}
In this section, we discuss the worst-case scenario for  solving MaxCut of $3$-regular graph using $p=1$ QAOA, and provide an example of a planar graph that can be solved efficiently.
For a $p=1$ QAOA, the approximation ratio $\alpha$ of a $3$-regular graph will depend on three subgraphs, $g_4,g_5,g_6$~\cite{Farhi2014,Wurtz2021}, where the subscripts represent the number of vertices included in the subgraph. In the worst-case, the entire graph consists solely of the subgraph $g_6$(see~\Cref{fig:G6}), and the local cut approximation ratio for $g_6$ in a $p=1$ QAOA is bounded by
\begin{equation}\label{eq:g_6def}
\begin{aligned}
 \alpha_{g_6} & = \frac{1-\langle +|e^{i\gamma_1 C_{g_6}} e^{i\beta B_{g_6}} Z_2Z_3 e^{-i\beta B_{g_6}} e^{-i\gamma_1 C_{g_6}}|+\rangle}{2c_{g_6}} \\
 & \leq 0.6924
 \end{aligned}
\end{equation}
where $c_{g_6}$ is the local maxcut and is equal to $1$ for the $g_6$, $C_{g_6} = Z_0Z_2+Z_1Z_2+Z_2Z_3+Z_3Z_4+Z_3Z_5$, $B_{g_6} = X_2+X_3$. Here, the equality is achieved when $\gamma=17.8^{\circ}, \beta = 22.5^{\circ}$. It is noted that our value of $\gamma$ differs from Ref.~\cite{Wurtz2021}, as we use a reduced $C_{g_6}$ without loss of generality. The approximation ratio for the entire graph is then bounded by $\alpha_{g_6}$, since the average cut is the same as each edge if contains one type of subgraph. 

It is known that the MaxCut of a planar graph can be solved in polynomial time~\cite{Barahona1983}. For example, we find such a planar graph by connecting a regular polygon with even number of edges to the same polygon vertex-by-vertex---see the square example in~\Cref{fig:G6}. If we group the odd and even vertices into separate sets, we find the graph's MaxCut for cutting all edges. Thus, discussing the solvability of planar graphs using QAOA is less significant, since polynomial-time solutions already exist.

\begin{figure}[h]
    \centering
    \input{tikz/g6.tikz}
    \hspace{1cm}
    \input{tikz/G6_8.tikz}
    \caption{The subgraph $g_6$ of the $3$-regular graph using $p=1$ QAOA(left panel); A worst-case of the $3$-regular planar graphs using $p=1$ QAOA(right panel). The dots are vertices with the number indexed underneath. The lines connected vertices are the edges.}\label{fig:G6}
\end{figure}

\section{The scaling of p for different statistic order of prepared initial state}\label{sec:app_p_impact}
\begin{figure}[h]
  \centering
    \includegraphics{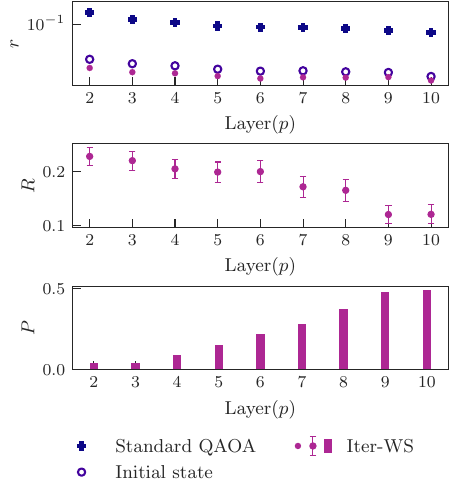}
    \caption{The $r$(a), $R$(b) and $P$(c) for the iterative warm-started quantum approximate optimisation applied over one iteration to $100$ randomly generated $3$-regular graphs with $12$ vertices each, as a function of the layer $p$ of the standard QAOA. The prepared initial state is a $15$-th ordered statistic state. The classical optimiser is DA with $\mathcal{I} = 5000$, $m = 8000$, and $M = 8000$.}\label{fig:p_list_15}
\end{figure}

This section extends the discussion of the influence of the layer $p$ in the standard QAOA on the stochastic state order $t$ when solving $3$-regular graphs. We observe that for $t=15$ (see~\Cref{fig:p_list_15}), the results show a similar plateaus-to-decrease trend for $R$ and a increasing convergence trend for $P$, as seen for $t=20$'s simulations (see ~\Cref{fig:p_list_20}).

\begin{figure}[h]
  \centering
    \includegraphics{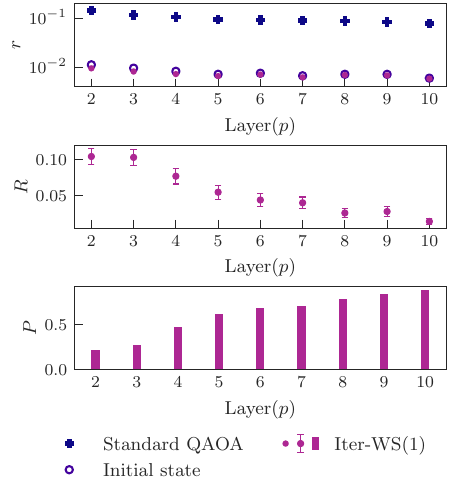}
    \caption{The $r$(a), $R$(b) and $P$(c) for the iterative warm-started quantum approximate optimisation applied over one iteration to $100$ randomly generated $3$-regular graphs with $12$ vertices each, as a function of the layer $p$ of the standard QAOA. The prepared initial state is a $5$-th ordered statistic state. The classical optimiser is DA with $\mathcal{I} = 5000$, $m = 8000$, and $M = 8000$.}\label{fig:p_list_5}
\end{figure}

However, when we reduce the order $t$ to $5$, both $R$ and the transition point of $R$ decrease, and $P$ generally increases (see~\Cref{fig:p_list_5}). The poorer performance might be due to the fact that a $t$-th ordered stochastic state will approximate a warm-started state prepared by the classical bit string. 

\end{document}

%% file: tikz/g6.tikz
\begin{tikzpicture}
    \coordinate (j) at (0,0);
    \coordinate (k) at (1,0);
    \coordinate (j1) at (-0.75,0.75);
    \coordinate (j2) at (-0.75,-0.75);
    \coordinate (k1) at (1.75,0.75);
    \coordinate (k2) at (1.75,-0.75);

    \draw[thick] (j) -- (j1);
    \draw[thick] (j) -- (j2);
    \draw[thick] (j) -- (k);
    \draw[thick] (k) -- (k1);
    \draw[thick] (k) -- (k2);

    \fill (j) circle (3.6pt);
    \fill (k) circle (3.6pt);
    \fill (j1) circle (3.6pt);
    \fill (j2) circle (3.6pt);
    \fill (k1) circle (3.6pt);
    \fill (k2) circle (3.6pt);

    \node[below=1.2mm] at (j1) {$0$};
    \node[below=1.2mm] at (j2) {$1$};
    \node[below=1.2mm] at (j) {$2$};
    \node[below=1.2mm] at (k) {$3$};
    \node[below=1.2mm] at (k1) {$4$};
    \node[below=1.2mm] at (k2) {$5$};
\end{tikzpicture}

%% file: tikz/G6_8.tikz
\begin{tikzpicture}[scale=0.6] 
    \coordinate (A) at (0, 0);
    \coordinate (B) at (5, 0);
    \coordinate (C) at (5, 5);
    \coordinate (D) at (0, 5);

    \coordinate (center) at (2.5, 2.5);  

    \coordinate (E) at (1.5, 1.5);  
    \coordinate (F) at (3.5, 1.5);  
    \coordinate (G) at (3.5, 3.5);  
    \coordinate (H) at (1.5, 3.5);  

    \draw[thick] (A) -- (B) -- (C) -- (D) -- cycle;

    \draw[thick] (E) -- (F) -- (G) -- (H) -- cycle;

    \draw[thick] (A) -- (E);
    \draw[thick] (B) -- (F);
    \draw[thick] (C) -- (G);
    \draw[thick] (D) -- (H);

    \fill (A) circle (6pt);
    \fill (B) circle (6pt);
    \fill (C) circle (6pt);
    \fill (D) circle (6pt);

    \fill (E) circle (6pt);
    \fill (F) circle (6pt);
    \fill (G) circle (6pt);
    \fill (H) circle (6pt);

    \node[below=1mm] at (E) {$1$};
    \node[below=1mm] at (F) {$2$};
    \node[above=1mm] at (G) {$3$};
    \node[above=1mm] at (H) {$4$};

    \node[below=2mm] at (A) {$5$};
    \node[below=2mm] at (B) {$6$};
    \node[above=2mm] at (C) {$7$};
    \node[above=2mm] at (D) {$8$};
\end{tikzpicture}

%% file: main.bbl
\begin{thebibliography}{10}

\bibitem{Cain2023}
Madelyn Cain, Edward Farhi, Sam Gutmann, Daniel Ranard, and Eugene Tang.
\newblock The qaoa gets stuck starting from a good classical string, 2023.

\bibitem{Quantum2020}
Google~AI Quantum, Collaborators*†, Frank Arute, Kunal Arya, Ryan Babbush, Dave Bacon, Joseph~C. Bardin, Rami Barends, Sergio Boixo, Michael Broughton, Bob~B. Buckley, David~A. Buell, Brian Burkett, Nicholas Bushnell, Yu~Chen, Zijun Chen, Benjamin Chiaro, Roberto Collins, William Courtney, Sean Demura, Andrew Dunsworth, Edward Farhi, Austin Fowler, Brooks Foxen, Craig Gidney, Marissa Giustina, Rob Graff, Steve Habegger, Matthew~P. Harrigan, Alan Ho, Sabrina Hong, Trent Huang, William~J. Huggins, Lev Ioffe, Sergei~V. Isakov, Evan Jeffrey, Zhang Jiang, Cody Jones, Dvir Kafri, Kostyantyn Kechedzhi, Julian Kelly, Seon Kim, Paul~V. Klimov, Alexander Korotkov, Fedor Kostritsa, David Landhuis, Pavel Laptev, Mike Lindmark, Erik Lucero, Orion Martin, John~M. Martinis, Jarrod~R. McClean, Matt McEwen, Anthony Megrant, Xiao Mi, Masoud Mohseni, Wojciech Mruczkiewicz, Josh Mutus, Ofer Naaman, Matthew Neeley, Charles Neill, Hartmut Neven, Murphy~Yuezhen Niu, Thomas~E. O’Brien, Eric Ostby, Andre Petukhov, Harald
  Putterman, Chris Quintana, Pedram Roushan, Nicholas~C. Rubin, Daniel Sank, Kevin~J. Satzinger, Vadim Smelyanskiy, Doug Strain, Kevin~J. Sung, Marco Szalay, Tyler~Y. Takeshita, Amit Vainsencher, Theodore White, Nathan Wiebe, Z.~Jamie Yao, Ping Yeh, and Adam Zalcman.
\newblock Hartree-fock on a superconducting qubit quantum computer.
\newblock {\em Science}, 369(6507):1084--1089, August 2020.

\bibitem{Grimsley2023}
Harper~R. Grimsley, George~S. Barron, Edwin Barnes, Sophia~E. Economou, and Nicholas~J. Mayhall.
\newblock Adaptive, problem-tailored variational quantum eigensolver mitigates rough parameter landscapes and barren plateaus.
\newblock {\em npj Quantum Information}, 9(1):19, March 2023.

\bibitem{Farhi2014}
Edward Farhi, Jeffrey Goldstone, and Sam Gutmann.
\newblock A quantum approximate optimization algorithm, November 2014.

\bibitem{Cook2020}
Jeremy Cook, Stephan Eidenbenz, and Andreas B{\"a}rtschi.
\newblock The quantum alternating operator ansatz on maximum k-vertex cover.
\newblock In {\em 2020 IEEE International Conference on Quantum Computing and Engineering (QCE)}, pages 83--92. IEEE, 2020.

\bibitem{Egger2021}
Daniel~J. Egger, Jakub Marecek, and Stefan Woerner.
\newblock Warm-starting quantum optimization.
\newblock {\em Quantum}, 5:479, June 2021.

\bibitem{Tate2023}
Reuben Tate, Jai Moondra, Bryan Gard, Greg Mohler, and Swati Gupta.
\newblock Warm-started {QAOA} with {C}ustom {M}ixers {P}rovably {C}onverges and {C}omputationally {B}eats {G}oemans-{W}illiamson's {M}ax-{C}ut at {L}ow {C}ircuit {D}epths.
\newblock {\em Quantum}, 7:1121, September 2023.

\bibitem{yuan2024quantifyingadvantagesapplyingquantum}
Haomu Yuan, Christopher~K. Long, Hugo~V. Lepage, and Crispin H.~W. Barnes.
\newblock Quantifying the advantages of applying quantum approximate algorithms to portfolio optimisation, 2024.

\bibitem{He2023}
Zichang He, Ruslan Shaydulin, Shouvanik Chakrabarti, Dylan Herman, Changhao Li, Yue Sun, and Marco Pistoia.
\newblock Alignment between initial state and mixer improves qaoa performance for constrained optimization.
\newblock {\em npj Quantum Information}, 9(1), November 2023.

\bibitem{Wang2018}
Zhihui Wang, Stuart Hadfield, Zhang Jiang, and Eleanor~G. Rieffel.
\newblock Quantum approximate optimization algorithm for maxcut: A fermionic view.
\newblock {\em Phys. Rev. A}, 97:022304, Feb 2018.

\bibitem{Shaydulin2024}
Ruslan Shaydulin, Changhao Li, Shouvanik Chakrabarti, Matthew DeCross, Dylan Herman, Niraj Kumar, Jeffrey Larson, Danylo Lykov, Pierre Minssen, Yue Sun, Yuri Alexeev, Joan~M. Dreiling, John~P. Gaebler, Thomas~M. Gatterman, Justin~A. Gerber, Kevin Gilmore, Dan Gresh, Nathan Hewitt, Chandler~V. Horst, Shaohan Hu, Jacob Johansen, Mitchell Matheny, Tanner Mengle, Michael Mills, Steven~A. Moses, Brian Neyenhuis, Peter Siegfried, Romina Yalovetzky, and Marco Pistoia.
\newblock Evidence of scaling advantage for the quantum approximate optimization algorithm on a classically intractable problem.
\newblock {\em Science Advances}, 10(22), May 2024.

\bibitem{Farhi2020}
Edward Farhi, David Gamarnik, and Sam Gutmann.
\newblock The quantum approximate optimization algorithm needs to see the whole graph: Worst case examples, 2020.

\bibitem{Wurtz2021}
Jonathan Wurtz and Peter Love.
\newblock Maxcut quantum approximate optimization algorithm performance guarantees for $p>1$.
\newblock {\em Phys. Rev. A}, 103:042612, Apr 2021.

\bibitem{Akshay2021}
V.~Akshay, H.~Philathong, I.~Zacharov, and J.~Biamonte.
\newblock Reachability {D}eficits in {Q}uantum {A}pproximate {O}ptimization of {G}raph {P}roblems.
\newblock {\em {Quantum}}, 5:532, August 2021.

\bibitem{Larkin2022}
Jason Larkin, Matías Jonsson, Daniel Justice, and Gian~Giacomo Guerreschi.
\newblock Evaluation of qaoa based on the approximation ratio of individual samples.
\newblock {\em Quantum Science and Technology}, 7(4):045014, aug 2022.

\bibitem{Tate2024}
Reuben Tate and Stephan Eidenbenz.
\newblock Theoretical approximation ratios for warm-started qaoa on 3-regular max-cut instances at depth $p=1$, 2024.

\bibitem{Goemans1995}
Michel~X. Goemans and David~P. Williamson.
\newblock Improved approximation algorithms for maximum cut and satisfiability problems using semidefinite programming.
\newblock {\em J. ACM}, 42(6):1115–1145, November 1995.

\bibitem{Halperin2004}
Eran Halperin, Dror Livnat, and Uri Zwick.
\newblock Max cut in cubic graphs.
\newblock {\em Journal of Algorithms}, 53(2):169--185, 2004.

\bibitem{Hodson2019}
Mark Hodson, Brendan Ruck, Hugh Ong, David Garvin, and Stefan Dulman.
\newblock Portfolio rebalancing experiments using the quantum alternating operator ansatz, November 2019.

\bibitem{Brandhofer2022}
Sebastian Brandhofer, Daniel Braun, Vanessa Dehn, Gerhard Hellstern, Matthias H{\"u}ls, Yanjun Ji, Ilia Polian, Amandeep~Singh Bhatia, and Thomas Wellens.
\newblock Benchmarking the performance of portfolio optimization with {{QAOA}}.
\newblock {\em Quantum Information Processing}, 22(1):25, December 2022.

\bibitem{Herman2022}
Dylan Herman, Cody Googin, Xiaoyuan Liu, Alexey Galda, Ilya Safro, Yue Sun, Marco Pistoia, and Yuri Alexeev.
\newblock A survey of quantum computing for finance, January 2022.

\bibitem{Chen2024}
Bingren Chen, Hanqing Wu, Haomu Yuan, Lei Wu, and Xin Li.
\newblock Quasi-binary encoding based quantum alternating operator ansatz, January 2024.

\bibitem{Ramacciotti2024}
Debora Ramacciotti, Andreea~I. Lefterovici, and Antonio~F. Rotundo.
\newblock Simple quantum algorithm to efficiently prepare sparse states.
\newblock {\em Phys. Rev. A}, 110:032609, Sep 2024.

\bibitem{Yordanov2020}
Yordan~S. Yordanov, David R.~M. Arvidsson-Shukur, and Crispin H.~W. Barnes.
\newblock Efficient quantum circuits for quantum computational chemistry.
\newblock {\em Phys. Rev. A}, 102:062612, Dec 2020.

\bibitem{Jordan1928}
P.~Jordan and E.~Wigner.
\newblock Über das paulische Äquivalenzverbot.
\newblock {\em Zeitschrift für Physik}, 47(9–10):631--651, September 1928.

\bibitem{Hadfield2019}
Stuart Hadfield, Zhihui Wang, Bryan O’Gorman, Eleanor~G. Rieffel, Davide Venturelli, and Rupak Biswas.
\newblock From the quantum approximate optimization algorithm to a quantum alternating operator ansatz.
\newblock {\em Algorithms}, 12(2), 2019.

\bibitem{Zhou2022}
Leo Zhou, Sheng-Tao Wang, Soonwon Choi, Hannes Pichler, and Mikhail~D. Lukin.
\newblock Quantum approximate optimization algorithm: Performance, mechanism, and implementation on near-term devices.
\newblock {\em Phys. Rev. X}, 10:021067, Jun 2020.

\bibitem{Harrigan2021}
Matthew~P. Harrigan, Kevin~J. Sung, Matthew Neeley, Kevin~J. Satzinger, Frank Arute, Kunal Arya, Juan Atalaya, Joseph~C. Bardin, Rami Barends, Sergio Boixo, Michael Broughton, Bob~B. Buckley, David~A. Buell, Brian Burkett, Nicholas Bushnell, Yu~Chen, Zijun Chen, Ben Chiaro, Roberto Collins, William Courtney, Sean Demura, Andrew Dunsworth, Daniel Eppens, Austin Fowler, Brooks Foxen, Craig Gidney, Marissa Giustina, Rob Graff, Steve Habegger, Alan Ho, Sabrina Hong, Trent Huang, L.~B. Ioffe, Sergei~V. Isakov, Evan Jeffrey, Zhang Jiang, Cody Jones, Dvir Kafri, Kostyantyn Kechedzhi, Julian Kelly, Seon Kim, Paul~V. Klimov, Alexander~N. Korotkov, Fedor Kostritsa, David Landhuis, Pavel Laptev, Mike Lindmark, Martin Leib, Orion Martin, John~M. Martinis, Jarrod~R. McClean, Matt McEwen, Anthony Megrant, Xiao Mi, Masoud Mohseni, Wojciech Mruczkiewicz, Josh Mutus, Ofer Naaman, Charles Neill, Florian Neukart, Murphy~Yuezhen Niu, Thomas~E. O’Brien, Bryan O’Gorman, Eric Ostby, Andre Petukhov, Harald Putterman, Chris
  Quintana, Pedram Roushan, Nicholas~C. Rubin, Daniel Sank, Andrea Skolik, Vadim Smelyanskiy, Doug Strain, Michael Streif, Marco Szalay, Amit Vainsencher, Theodore White, Z.~Jamie Yao, Ping Yeh, Adam Zalcman, Leo Zhou, Hartmut Neven, Dave Bacon, Erik Lucero, Edward Farhi, and Ryan Babbush.
\newblock Quantum approximate optimization of non-planar graph problems on a planar superconducting processor.
\newblock {\em Nature Physics}, 17(3):332--336, February 2021.

\bibitem{Barahona1983}
Francisco Barahona.
\newblock The max-cut problem on graphs not contractible to k5.
\newblock {\em Operations Research Letters}, 2(3):107--111, 1983.

\end{thebibliography}
